\documentclass[fleqn,usenatbib]{mnras}

\usepackage{newtxtext,newtxmath}

\usepackage[T1]{fontenc}

\DeclareRobustCommand{\VAN}[3]{#2}
\let\VANthebibliography\thebibliography
\def\thebibliography{\DeclareRobustCommand{\VAN}[3]{##3}\VANthebibliography}

\usepackage{booktabs}
\usepackage{tikz}
\usetikzlibrary{positioning,arrows.meta,fit,calc,backgrounds}
\usetikzlibrary{decorations.pathreplacing,fit,calc,positioning,arrows.meta}
\usepackage{verbatim}
\usepackage{amsmath}

\usepackage{graphicx}	
\usepackage{amsmath}	
\usepackage{orcidlink}

\title[SBI population modelling]{Simulation-based inference for AGN jet population modelling: Towards more robust comparisons of black hole jet speeds}

\author[Lilje et al.]{
Clara Lilje$^{\orcidlink{0009-0004-1555-5356}}$,$^{1}$\thanks{E-mail: Clara.Lilje@physics.ox.ac.uk}
James H. Matthews$^{\orcidlink{0000-0002-5654-2744}}$,$^{1}$ 
and Rob Fender$^{\orcidlink{0000-0002-3493-7737}}$,$^{1,2}$
\\
$^{1}$Department of Physics, Astrophysics, University of Oxford, Denys Wilkinson Building, Keble Road, Oxford, OX1 3RH, UK.\\
$^{2}$Department of Astronomy, University of Cape Town, Private Bag X3, Rondebosch, 7701, South Africa.
}

\date{\today}

\pubyear{\the\year{}}

\begin{document}
\label{firstpage}
\pagerange{\pageref{firstpage}--\pageref{lastpage}}
\maketitle

\begin{abstract}
We present the most complete modelling of the MOJAVE 1.5 Jansky Quarter Century active galactic nuclei (AGN) jet population, using likelihood-free simulation-based inference. Due to the complex impact of a flux-limit on observed AGN data sets, careful modelling of the parent population is required. In particular, when observing and fitting to multiple data distributions likelihoods become non-intuitive. Parameter degeneracies further complicate the problem and make it suitable for likelihood-free, simulation-based inference. This method relies on a normalising flow learning the likelihood surface or posteriors directly. We extensively validate the flow to show that previous parameter estimates for the AGN jet speed distributions underestimated parameter errors significantly and do not capture the non-gaussianity of the parameter posteriors. The new results enable a better statistical comparison to other AGN population studies, but also a more accurate comparison of supermassive black hole jets  with their lower mass counterparts, X-ray binaries (XRB). We find that the AGN follow a Lorentz factor distribution of the shape $N(\Gamma) \propto \Gamma^b$ with $b= -1.32_{-0.19}^{+0.20}$. This slope is consistent with the XRB Lorentz factor distribution at $2\sigma$.
Simulation-based inference as a method is generally well-suited to many astrophysical problems, and this paper shows the convenient applicability of this methodology to parent population studies of jetted AGN with multiple observables specifically.
\end{abstract}

\begin{keywords}
galaxies: jets -- X-rays: binaries -- ISM: jets and outflows -- galaxies: active -- black hole physics
\end{keywords}



\section{Introduction}
Astrophysical jets are produced by some of the most compact objects in our Universe. They propagate over large distances but originate from the very smallest scales and link closely to the accretion process \citep[e.g.][]{fender_towards_2004}. 
In particular,  black holes (BH) are ideal laboratories as they are intrinsically simple objects, only governed by a few parameters, such as mass and spin. In addition, while they span many orders of magnitude, especially across mass, they remain uniquely scalable, as seen in the fundamental plane of black hole activity \citep{merloni_fundamental_2003,falcke_scheme_2004,plotkin_using_2012}. As they probe different  time scales in jet formation and propagation, populations at both ends of the mass spectrum can be used to gain insight into complementary aspects of jet physics.

Parent population studies can reveal the properties of all sources in an object class, even if we are only able to observe a biased sample of these sources. Jetted BHs are a class of sources that is usually observed in an extremely biased sample as relativistic boosting severely brightens sources which are fast and aligned with respect to the line of sight. Off-axis sources, on the other hand, appear slower due to the apparent speed flattening off at larger Lorentz factors at high inclinations \citep{lilje_kinematics_2025}.
However, if we are able to appropriately model which distributions of source properties we would expect and understand which biases we need to account for, it is possible to infer the parameters of the parent population distributions.
Although there are simple, established tools to infer parameter distributions, such as Anderson-Darling \citep{anderson_asymptotic_1952} or $\chi^2$ tests, these methods struggle when it comes to fitting multiple observables simultaneously and inferring appropriate posteriors for each parameter. Previous results using these methods have already shown intriguing agreement between jet populations across the BH mass range \citep{lilje_kinematics_2025}, but the lack of full posteriors has limited the statistical significance of the comparison. 
Therefore, introducing novel methods, such as simulation-based inference, is extremely powerful in fully constraining the parent population properties of biased samples.
Astrophysics as a field has become very simulation-driven, we are able to simulate non-linear process affecting our data more and more efficiently. At the same time, our data sets are becoming more and more complex, which can make likelihoods intractable. Simulation-based inference (SBI) is the ideal tool for inference on such problems. It is already commonly used in cosmology, but can easily be implemented in other areas of astrophysics \citep{cranmer_frontier_2020}. As implementations have become much more accessible, due to code packages which provide the machine learning infrastructure as an extremely flexible black box.
Furthermore, SBI methodologies are starting to be recognised for their flexibility specifically for population studies as shown by \cite{gerolymatou_population_2026}.

In this work we apply these novel SBI methods to a well-studied sample to confirm and extend previous analysis. We confirm that the simulation-based inference performs extremely well to construct posteriors for each parameter. As a result, we are able to compare the Lorentz factor distribution of AGN to XRBs and conclude that the slopes of both populations are consistent at $2\sigma$, which could indicate common jet physics across the mass range.
\section{Methods}
\label{sec:methods}
\begin{table*}
    \centering
    \begin{tabular}{cccc}
        \toprule
        Jet Property & Distribution & Fixed parameters & Free parameter range\\ \toprule
         Lorentz factor & $N(\Gamma)d\Gamma \propto \Gamma^b$ & \vtop{\hbox{\strut $\Gamma_{\text{min}} = 1.25$}\hbox{\strut  $\Gamma_{\text{max}} = 50$}}   & $-3.2 \leq b \leq -0.2$\\ \midrule
         Luminosity function & \vtop{\hbox{\strut $\Phi(L,z) \propto \Phi(L/e(z))$}\hbox{\strut  $e(z) = (1+z)^k e^{(z/\eta)}$}\hbox{\strut  $\Phi(L/e(z=0)) \propto L^\gamma$}} & \vtop{\hbox{\strut $L_{\text{min}} = 10^{24} \text{W Hz}^{-1}$}\hbox{\strut  $L_{\text{max}} = 10^{31} \text{W Hz}^{-1}$}} & \vtop{\hbox{\strut $-0.65 \leq \eta \leq -0.25$}\hbox{\strut  $4.5 \leq k \leq 9.5$}\hbox{\strut  $-3.6\leq \gamma \leq -2.4$}} \\ \midrule
         Beamed Luminosity & $P=L\delta^p$ & $p=2-\alpha$ &$-0.7 \leq \alpha \leq 0.5 $ \\ \midrule
         Viewing angle & $p(\theta)d\theta=\sin\theta$ & \vtop{\hbox{\strut $\theta_{\text{min}} = 0\textdegree$}\hbox{\strut  $\theta_{\text{max}} = 90\textdegree$}} & \\ \bottomrule
    \end{tabular}
    \caption{Distributions of Monte Carlo simulation parameters for the FSRQ sample as taken from \protect\cite{lister_mojave_2019}. The parameter ranges were adapted to improve posterior convergence and enable broader comparison to other samples.}
    \label{tab:parentdistr}
\end{table*}
\subsection{Parent Population modelling}
\label{sec:popmodel}
To investigate the intrinsic Lorentz factor distribution of AGN, we employ Monte Carlo parent population modelling based on the work by \cite{lister_mojave_2019}. We adopt the same form of parent population distributions, summarised in Table \ref{tab:parentdistr}, but we widen the explored parameter ranges in some cases to enable smoother posteriors.
In summary, the population is modelled with an isotropic inclination distribution in 3D space, which results in a viewing angle distribution of $p(\theta)d\theta =\sin\theta$. The Lorentz factors $\Gamma$ are modelled with a power law with index $b$. Similarly, the base distribution for the luminosity $L$ is also a power law with index $\gamma$ at redshift $z=0$. The redshift evolution is included in the luminosity function as a pure luminosity evolution with a constant comoving density with redshift. For further details, we refer the reader to \cite{lister_mojave_2019}.

While the functions describing the parent population themselves remain the same as in \cite{lister_mojave_2019}, we sample the priors more smoothly and fine-grained. This was possible by optimising the simulation process with \texttt{numba-jit} implementation \citep{lam_numba_2015}. We extend some of the priors to enable better posterior coverage. Due to computational limits and storage constraints, we limit the parameter precision to two decimal places. In most cases this is much below the posterior width and the posteriors remain smooth, therefore this truncation does not significantly impact the result. 

We sample all distributions by obtaining the inverse of the cumulative density function of each jet property. Each jet source is assigned a Lorentz factor $\Gamma$, intrinsic luminosity $L$, redshift $z$ and viewing angle $\theta$, which are used to calculate a beamed luminosity $P$ and flux $S$. The simulated jets are only added to the "observed" sample if their flux exceeds $S_j > 1.5 \rm Jy$. This implies that there is a much larger parent population generated than the "observed" sample, which is then compared to the observables of the MOJAVE 1.5 Jy QC FSRQ sample. 

\subsection{Simulation-based inference}
\label{sec:SBI}
To obtain the posteriors for each parameter, we utilise simulation-based inference (SBI). 
SBI is an umbrella term for performing inference on data for which the likelihood is intractable, but can be well-approximated via forward simulation. This method is also sometimes known as likelihood-free inference, but as \cite{cranmer_frontier_2020} point out, this is a bit of a misnomer, as the goal of SBI is usually to approximate this likelihood. 
We recommend \cite{cranmer_frontier_2020} and \cite{deistler_simulation-based_2025} for more detailed reviews on applied SBI in research.

Generally, SBI relies on a simulator capable of producing synthetic data sets similar to observed data based on a set of parameters drawn from known priors. Then, a machine learning (ML) inference network is trained on the sets of parameters and simulation outputs. The network learns which sets of parameters generate outputs which are close in simulation space. This means that the network learns to approximate the likelihood surface without the requirement for an analytical description of the likelihood. Once the inference network is trained, it can be evaluated on the observed data to obtain posteriors. This specific flavour of SBI is called neural posterior estimation (NPE). It is also possible to instead train the network to output the likelihood or likelihood ratio, known as neural likelihood or neural ratio estimation respectively. NPE has the advantage of performing very fast inference immediately as it is not necessary to sample the likelihood surface itself, and is sufficient for our problem.

We use uninformative, uniform priors for each parameter. Each set of parameters is used to generate a set of AGN with 4 simulated observables each. This has the same number of synthetic AGN as the observed distributions. We choose to compare to the sample chosen in \cite{lister_mojave_2019}, which consists of $174$ flat spectrum radio quasars (FSRQs). As these sources display superluminal jet components, they are the ideal class of objects to determine jet speed distributions. We limit the analysis to a sample of FSRQs as it is possible that BL Lacs follow different parent population distribution \citep{urry_unified_1995} as they may represent a different AGN state. For the SBI inference we simulate a data set of shape $[N_{\rm sim},174,4]$ and a parameter data set of shape $[N_{\rm sim}, 4]$.

\subsubsection{Data pre-processing}
The machine-learning algorithm performs best on data scaled between 0 and 1, that populate the domain roughly uniformly, rather than being concentrated in part of the range as for example with logarithmic values. Therefore we rescale beamed luminosity as $L_{\rm log} = \log_{10}(L_{\rm beamed})$ and then pre-process all observables with a \texttt{sklearn.preprocessing.MinMaxScaler} \citep{pedregosa_scikit-learn_2011}, which scales to values between 0 and 1. The scaler is fitted to a single simulation and then reused for all training, testing and observed data. Parameters are also rescaled to be between 0 and 1 before training. The training data is then split into a $90-10$ training-validation split. 

Additionally, the data in the FSRQ sub-sample of the MOJAVE 1.5 Jy QC survey has some missing apparent speed $\beta_{\rm app}$ values. To condition the flow to account for this, we randomly choose between $10 \% -50 \%$ of apparent speed values in the simulated data to be missing. We replace them with a value of $-1$ as this is far outside the learned distribution. Additionally, we add a fifth dimension to our data, which is a "missing" flag of $0$ or $1$, when the data is missing. This method was shown to be most effective in dealing with gaps in the data \citep{wang_missing_2024}.

\subsubsection{Flow structure}
We choose a relatively standard structure for the inference back end of our SBI pipeline. We also ensure to use encoders which are appropriate for the structure and type of data we are using. The architecture choice is then also validated through extensive hyperparameter searches and calibration testing.

We choose to perform the inference with a neural spline flow (NSF) \citep{durkan_neural_2019} implemented in the \texttt{sbi} python package. These flows are flexible and fully invertible while also performing density estimation at a comparable accuracy to state of the art models. The posterior estimator is usually best at dealing only with one-dimensional inputs and therefore it is advisable to encode the multi-dimensional data into a context vector to pass onto the NSF for training \citep{radev_bayesflow_2020}. This ensures that high-dimensional information in the data does not get lost by flattening the data or taking summary statistics to represent the population. First we encode a per-object feature representation with a fully-connected multi-layer neural network on each object independently with no mixing. This is the obvious choice for our model as our data set consists of independent objects which however have the same 5 observables describing each of them. By using shared weights across the each object, we allow the flow to learn that each observable should encode similar information for each object. 
Then we use permutation-invariant embedding as a population-level encoder. The permutation-invariance is an appropriate choice here as the order of objects in the observed sample should not make a difference to the output of the model. Lastly, we input this context vector along with the normalised parameters into the \texttt{sbi} posterior neural network. The network is trained to minimise validation loss using the atomic cross-entropy loss defined in \cite{greenberg_automatic_2019} and implemented in the NPE network functions in the \texttt{sbi} package. 

To ensure best performance we optimise the network architecture using \texttt{optuna} \citep{akiba_optuna_2019}. The different architectures are all optimised for best validation loss, but also evaluated on a held-out calibration test set. 
Posterior calibration is assessed using three complementary diagnostics. Simulation-based calibration (SBC) tests whether posterior rank statistics are uniformly distributed, with per-dimension Kalmogorov-Smirnov (KS) statistics reported as mean and worst-case values \citep{talts_validating_2020}. Empirical coverage is evaluated at the $90\%$ credible interval level, with deviations from the nominal coverage and normalised posterior-mean RMSE reported across parameter dimensions. Finally, TARP (Tests of Accuracy with Random Points) provides a global expected coverage probability test, yielding an area-to-diagonal statistic and an associated KS p-value \citep{lemos_sampling-based_2023}. All calibration scores were recorded and any trials which did not meet thresholds for the calibration metrics were discarded. We performed multiple optimisation runs and narrowed the search space guided by calibration and loss values. This analysis showed that NSF-based posterior inference performed better than masked autoregressive flows (MAF)-based posterior networks.

We trained 10 seeds of the three best performing trials and then chose the hyperparameter trial which had the best calibration metrics and consistency across the ten seeds. We chose the production seed shown in the posterior plots as the seed whose normalised metric vector has minimum Euclidean distance to the group mean. We show the calibration metrics in the appendix in Table \ref{tab:calib-metrics}. These calibration metrics are sampled with the same numbers of posterior samples for each seed. We optimised the number of posterior samples to give a good indication of calibration quality while retaining easy computability. The SBC KS statistic is evaluated on 500 simulations, giving a sampling resolution of approximately $1/\sqrt{500} \approx 0.045$, while the TARP KS statistic is evaluated on 300 simulations, giving a resolution of approximately $1/\sqrt{300} \approx 0.058$, meaning differences in KS statistics smaller than roughly $0.04 - 0.06$ are not meaningfully distinguishable from sampling noise at these sample sizes. We later run more tests with larger samples on the production seeds. These show that the KS and TARP values remain consistent at higher resolution within the standard deviation.

\begin{figure*}
    \centering
    \includegraphics[width=0.8\linewidth]{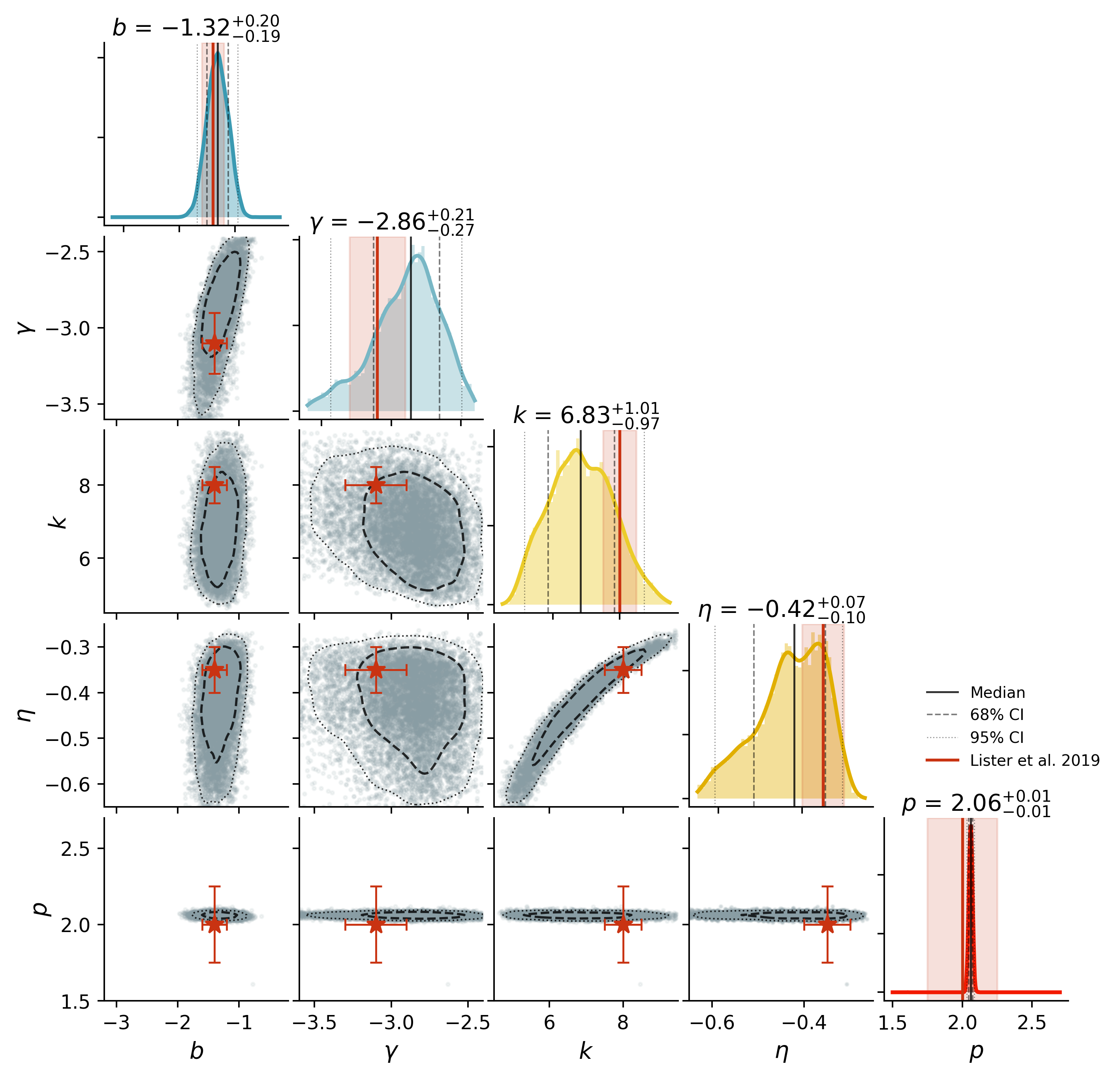}
    \caption{The parameter posteriors from the simulation-based inference model for the FSRQ sub-sample of the MOJAVE 1.5 Jy Quarter century AGN parent population. The red stars indicate the best fit values from \protect\cite{lister_mojave_2019}. The uncertainties quoted for each parameter are at the $1\sigma$ level.}
    \label{fig:corner_plot}
\end{figure*}
\section{Results}
\label{sec:Results}
Firstly, we train the flow to perform NPE, so the full end-to-end regression is performed in one evaluation of the flow \citep{greenberg_automatic_2019}. This is the fastest way to get the full posterior and is sufficient for our purposes. The training of the production seed converges after 364 epochs according to our patience criterion of no improvement for 50 epochs. The validation loss is consistently lower than the training loss which indicates that the model is not over-fitting. 
As discussed in the previous section, we first perform a general calibration analysis of the flow and hyperparameters to assess general performance. As the flow passes these tests to more than satisfactory standards, we move on to applying the model to real data.

In Figure \ref{fig:corner_plot} we show the posteriors obtained when applying the model to the FSRQ sub-sample of the MOJAVE 1.5 Jy QC sample. The posteriors appear well-converged and are not running up against prior boundaries. For reference we show the best fit values obtained by \cite{lister_mojave_2019}. The uncertainties for these literature values are simply the spacing between adjacent grid points which were analysed with the AD test. Within these uncertainties, it is obvious that the posteriors obtained in this work are consistent. In all cases the agreement is within $1 \sigma$, but most posteriors obtained using \texttt{sbi} are much broader. Some posteriors also exhibit significant asymmetry. To assess whether this posterior behaviour is reliable, we perform further convergence tests. The flow scores very well in the global and local calibration tests discussed in the previous section. These tests are the standard calibration tests and easily implemented with the \texttt{sbi} library. They are however exclusively simulation-based calibration, so they can only tell us about how well the flow performs on simulations generated with the same framework it is trained on. To ensure that the data we are providing to the flow is actually within this learned space, we also run a misspecification test using maximum mean discrepancy (MMD) \citep{schmitt_detecting_2021}. Using the FSRQ MOJAVE data set, we find a p-value of $0.994$, which cannot reject the null hypothesis that the observation was generated by the simulator. This strongly disfavours misspecification. In practice, this also means that the simulator we are using to generate the training data is a good proxy for the observed distribution of AGN jets. Additionally, we also perform a local classifier two sample tests (L-C2ST) which gives a necessary and sufficient condition for the validity of the algorithm \citep{linhart_l-c2st_2023}. The test evaluates the difference between the true and the estimated posterior distribution. With a p-value of $0.82$ we can once again say with $95\%$ confidence that the estimator is valid for our observation.

When investigating the correlation of the probability integral transforms of all parameter combinations, it becomes apparent that there is a strong correlation between $k$ and $\eta$. This is expected from previous results \citep{lister_mojave_2019}. As both parameters determine the redshift evolution of the sample, it becomes clear that there is a degeneracy in the model here that is not possible to disentangle. This degeneracy is therefore intrinsic and not a modelling failure of the flow. We can also see the impact of this effect in Figure \ref{fig:corner_plot}, where the 2D posterior of $k$ and $\eta$ shows a strong degeneracy. The SBI framework can capture this complexity of the model much better than AD tests. Additionally, we see that taking into account all distributions simultaneously leads to further posterior complexity, such as the slight degeneracy between $\gamma$ (Luminosity function exponent) and $b$ (Lorentz factor distribution exponent). 

Lastly, it is important to address the very well constrained posterior of $p$ (beaming exponent). The flow is extremely confident in its estimation of $p$, giving an uncertainty down to the level of prior grid size. While this may seem concerning, it is not unexpected as the beaming exponent is one of the most easily learnable parameters, due to its direct mapping onto an observable - the beamed luminosity. This result re-emphasises that the choice by \cite{lister_mojave_2019} of fixing $p=2.0$ is appropriate for this data set. If we do this and only perform inference for the remaining 4 parameters the results stay very consistent, retaining the trends in posterior shape. The median values also stay consistent and are always within $1\sigma$. However, while performing the initial infrastructure tests, there were many flow set-ups which were consistently over-confident in $p$, this was easily identified in the per-parameter SBC rank plots. Where the plot for $p$ deviated strongly from the ideal uniform scenario. The fact that these plots, shown in the Appendix figure \ref{fig:sbc-rank-plots} do not show any strong deviations from the ideal line further confirms that the behaviour seen in the posterior for $p$ is indeed just a well-identified parameter and not severe over-confidence. 

\section{Discussion}
\label{sec:Discuss}

\begin{figure*}
    \centering
    \includegraphics[width=0.6\linewidth]{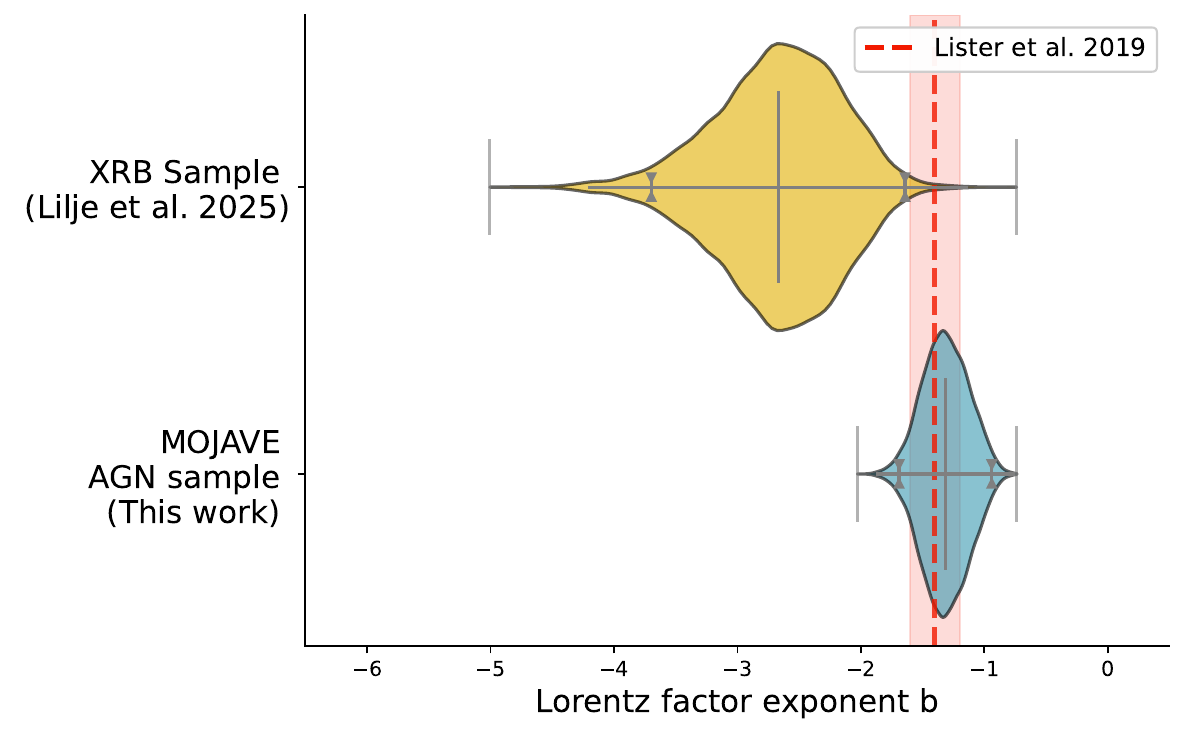}
    \caption{Comparison of posterior for the Lorentz factor exponent $b$, if $N(\Gamma) \propto \Gamma^b$. We show both the posteriors for the XRB large-scale jet population as taken from \protect\cite{lilje_kinematics_2025} and the posterior for the FSRQ sub-sample of the MOJAVE AGN sample obtained by SBI in this work. A vertical line marks the mean of the population and the arrow markers indicate the $2\sigma$ intervals of both posteriors. As a reference we also show the previous constraints from \protect\cite{lister_mojave_2019} obtained from the same sample.}
    \label{fig:xrb_comp}
\end{figure*}
\subsection{AGN jet parent population study comparison}
As discussed in the previous section the results shown in this work are consistent with literature results obtained from the same sample \citep{lister_mojave_2019}. 
As there are multiple flux-limited AGN samples selected at different wavelengths present in the literature, it is interesting to compare how the parameters found to describe the general parent population vary across these samples. Ideal other candidates are the BASS sample of Swift-BAT selected AGN \citep{marcotulli_bass_2022} and the Fermi-LAT selected AGN in \cite{ajello_luminosity_2012}. However there are some complications where despite best efforts the selection in different wavelength bands likely slightly biases the parent population estimates. X-ray or $\gamma$-ray selections likely target different parts of the jet than the radio emission which is produced further away from the central engine. In addition, different distribution cut-offs are used. For example, in \cite{ajello_luminosity_2012} it is shown that the Lorentz factor distribution of AGN follow $N(\Gamma) \propto \Gamma^{b}$ with $b=-2.03 \pm 0.70$ compared to our value of $b= -1.32_{-0.19}^{+0.20}$ which is consistent, but while we use $\Gamma_{\rm min } = 1.25$ and $\Gamma_{\rm max} = 50$, they use $\Gamma_{\rm min } = 5$ and $\Gamma_{\rm max} = 40$. In particular the much higher minimum cut-off is likely somewhat degenerate with the inference of the shape of the power-law. Their luminosity function exponent $\Phi(L/e(z=0)) \propto L^\gamma$ is $\gamma = -3.04\pm 0.08$ which is also consistent with our $\gamma =-2.86^{+0.20}_{-0.26}$. So, even with slight modelling discrepancies the overall agreement between the parent population distributions is good. 

For the BASS sample we find similar agreement where their equivalent $\gamma =-2.72 \pm 0.05$, the exponent $b$ is hugely dependant on their choice of beaming exponent $p$, where $p=7$ gives $b=-1.95\pm1.53$ and $p=5$ gives $b=-3.33\pm1.3$. Both of these solutions are degenerate. In both of these samples only the luminosity function of observed sources is fit, so it is possible that the extra information obtained through proper motion and redshift measurements adds more context to the fit.
 
There are also indications that there may be a population difference in sources selected in different wavelength bands.
\cite{lister_connection_2009} indicate that \textit{Swift}-BAT selected sources (BASS) tend to have faster estimated jet speeds than non-BAT selected AGN. \textit{Swift} light curves are much more sparsely sampled than \textit{Fermi}-LAT, which may also contribute to differences in the observed samples \citep{hovatta_relativistic_2020}. 

\subsection{Comparison to XRB jet speeds}
Lastly, we use these new constraints on the FSRQ jet Lorentz factor distribution to compare to existing estimates of the XRB population. Quantifying the agreement between these populations can enable us to draw conclusions about the fundamental properties of jets across the mass range, as discussed by \cite{lilje_kinematics_2025}. In Figure \ref{fig:xrb_comp}, we show the posteriors for the Lorentz factor distribution exponent $b$ for both the AGN sub-sample of FSRQs and XRBs, we can see that due to the width of the posterior distribution, the populations are in agreement at $2\sigma$ level. 

It is important to briefly discuss the choice of XRB sample. We compare the FSRQ jets here only to the large-scale relativistic jets or "ejecta" in XRBs, which were first observed to display the superluminal motions reminiscent of blazars by \cite{mirabel_superluminal_1994}. These jets are launched at the transition between the hard and soft spectral state in the XRB outburst cycle \citep{fender_towards_2004} and are distinct from the steady, continuous jets seen in the XRB hard state. As argued by \cite{lilje_kinematics_2025} the large-scale jets are the most comparable to FSRQs; the novel radio kinematics classifications presented by \cite{hervet_innovative_2016} show that FSRQs are very consistently classified separately to BL Lacs due to their super-luminal radio components. Indeed, in an approximate sense, the BL Lac and FSRQ radio phenomenology could be analogous to the two XRB jet phenomena (hard state steady jets and transient ejections, respectively). While there is some common behaviour between FSRQ jets and XRB ejecta, the true degree to which they are analogous phenomena is not completely clear (see e.g. discussion in \cite{moravec_radio_2022} and references therein). A comparison of their speeds is interesting regardless, given that it probes jet physics on vastly different mass-scales.

In most cases only a single large-scale ejection is observed to be launched at the state transition, but there are in fact examples of XRBs, especially GRS 1915+105 rapidly and repeatedly cycling through states \citep{belloni_model-independent_2000}, which are connected with rapid, repeated jet ejections \citep{fender_rapid_1997,klein-wolt_hard_2002}. Slowed to the supermassive black hole fundamental timescales, these systems would look quite similar as the multiple components seen in FSRQs. In addition, recent work by \cite{wood_ejection_2025} and \cite{hughes_spectropolarimetric_2026} shows that it is in fact likely that XRBs eject multiple radio "blobs" at the VLBI scale, with only a few remaining visible over the larger scale associated with the jets in the XRB sample compared to here.

The marginal $2\sigma$ discrepancy between the inferred values of the Lorentz factor distribution exponent $b$ can probably still be understood in terms of the caveats described in \cite{lilje_kinematics_2025}. In summary of those arguments, the fact that only the maximal apparent speed is considered in the AGN analysis may bias towards flatter power laws. It is unclear whether these maximal apparent speeds would in fact correspond to the apparent speeds of the relativistic "blobs" in XRBs which we see survive until the largest scales. Advances in VLBI imaging, such as the time-dependent visibility modelling presented by \cite{wood_ejection_2025} will certainly help to constrain the population of ejections within the same source in XRBs, just as it is possible with AGN and specifically blazars. It should be noted however that some sources, such as GRS 1915+105 show the same apparent speed over many ejections (between 1994-2018; see discussion in \citealt{miller-jones_deceleration_2006}; since 2018 the system's behaviour has changed considerably along with its obscuration state and the jets now seem to be slower and changing direction, see e.g. \cite{jiang_large_2026}).
In addition, we are probing much smaller scales in gravitational radii in the AGN; this may not be a significant problem, because the XRB jets appear to decelerate late in their evolution, when interacting with the interstellar medium \citep[e.g.][]{carotenuto_black_2021,savard_relativistic_2025}. 

It is understood that different wavelengths probe the jet at different distances from the central engine \citep{hovatta_relativistic_2020}.
Therefore, one may expect a change in parent Lorentz factor distribution for samples selected at different wavelengths, however using FSRQ samples selected in radio, x-rays and $\gamma$-rays as shown in \cite{lilje_kinematics_2025} seems to indicate that there appears to be consistency in jet Lorentz factor distribution even when probing different distances along the jet. Therefore the fact that we are probing different jet scales in XRBs and FSRQs may not have a large impact on the difference in the Lorentz factor distribution exponent $b$.

It could also be considered how jet structure impacts these results; should jets have a spine-sheath structure (e.g. \citealt{hardee_grmhdrmhd_2007,mizuno_three-dimensional_2007,qian_jet_2018,dihingia_thin_2024}) we would expect to see the sheath emission in XRBs and the spine emission in blazars due to boosting effects. As discussed by \cite{lilje_kinematics_2025} these effects may be able to account for some of the discrepancy between the sample of XRBs and FSRQs.
Regardless, these improved constraints enable us to say with confidence that the two jet populations are in agreement at the $2 \sigma$ level, which hints towards common jet physics and re-emphasizes that XRB jets may be just as relativistic as those in supermassive BHs \citep[see also ][]{zhang_jets_2025}. 

\section{Summary and Conclusions}
Our novel analysis of the MOJAVE FSRQ jet parent population extends previous studies to find improved posteriors for the parent population parameters. In particular, including multiple distributions of observables simultaneously into the fit and properly accounting for degeneracies shows that the parameter uncertainties are much broader and more asymmetric than shown in previous works. 
This effect actually leads to \emph{better} agreement with other AGN jet population studies performed on samples selected in other wavelength bands, such as the BASS or \textit{Fermi}-LAT selected AGN. 

In addition, we are able to refine the degree of agreement with the XRB jet population at the other end of the mass range. We find that the slopes of the Lorentz factor distribution of both populations is consistent within $2\sigma$. While there are some further caveats to this comparison, this study gives interesting hints towards similar jet physics in BHs across the mass range. 

These tests of jet physics are only possible with statistically sound methods that are able to take into account the non-linear impact of a flux limit on multiple observables, especially in the context of relativistic beaming.
We show that the SBI framework is ideal for these simulate-able problems of parameter inference. We do, however, caution readers that proper calibration tests of all ML infrastructure are key to ensure the reliability of results as over-fitting and over-confidence are common pathologies of these flows. Many of these problems can be improved with the correct hyperparameter choices.

\section*{Acknowledgements}
The authors acknowledge valuable conversations with Brian Rogers and Matthew L. Lister.
The authors thank the anonymous referee for their helpful report and insightful comments, which improved the clarity of the paper.
CL acknowledges support from the ERC.

RPF acknowledges support from UKRI, The ERC and The Hintze Family Charitable Foundation. 

JM acknowledges funding from a Royal Society University Research Fellowship (URF$\backslash$R1$\backslash$221062).
\section*{Data Availability}

The BHXRB data can be accessed as part of the data published by \cite{fender_speeds_2025}. And the AGN data used can be found in \cite{lister_mojave_2019} and \cite{homan_mojave_2021} where the data is available in machine-readable format. 


\makeatletter
\let\oldurl\url
\renewcommand{\url}[1]{}
\makeatother

\bibliographystyle{mnras}
\bibliography{20260905_bib} 


\appendix
\renewcommand{\thefigure}{A\arabic{figure}}
\renewcommand{\thetable}{A\arabic{table}}

\begin{table*}
\label{tab:modelStructure}
\centering
\begin{tabular}{l l l l}
\toprule
\textbf{Component} & \textbf{Type / Description} & \textbf{Output Dimensions} & \textbf{Hyperparameters} \\
\midrule
\textbf{Single Object Encoding} &
\vtop{\hbox{\strut Fully connected network (MLP)} \hbox{\strut with ReLU activations} \hbox{\strut applied per object}}
& $(B,\;174,\;190)$
& \vtop{\hbox{\strut Hidden width $= 210$,} \hbox{\strut Depth $= 5$}} \\

\textbf{Population-wide Encoder} &
\vtop{\hbox{\strut Permutation-invariant aggregation:} \hbox{\strut mean pooling over the per-object} \hbox{\strut embeddings, followed by a fully} \hbox{\strut connected network with ReLU}}
& $(B,\;64)$
& \vtop{\hbox{\strut Hidden dimensions $= 128$,}  \hbox{\strut Depth $= 2$}} \\

\textbf{Conditional Flow (NSF)} &
Stack of neural spline coupling layers
& $(B,\;4)$
& \vtop{\hbox{\strut Number of transforms $= 8$,}  \hbox{\strut Hidden features $= 64$}}\\

\textbf{Base Distribution} &
Standard Normal $\mathcal{N}(\mathbf{0},\mathbf{I})$
& --
& -- \\

\textbf{Training Parameters} &
\vtop{\hbox{\strut SBI defaults: Adam optimiser, early stopping,} \hbox{\strut gradient clipping (max norm 5.0)}}
& --
& \vtop{\hbox{\strut Learning rate $ = 1.14\times10^{-4}$,} \hbox{\strut Batch size $= 128$,} \hbox{\strut Max epochs $= 500$,} \hbox{\strut Patience $= 50$}} \\
\bottomrule
\end{tabular}
\caption{Architecture and hyperparameters of the conditional normalising flow model.
The model predicts the conditional density $p(\boldsymbol{\theta}\,|\,\mathbf{x})$,
where $\boldsymbol{\theta}$ are the physical parameters and $\mathbf{x}$ the observed features
(174 objects $\times$ 5 features per simulation). The production model was selected as the per-metric
median seed across 10 independent training runs.}
\end{table*}

\begin{table*}
    \centering
    \begin{tabular}{cccc}
    \toprule    
        Metric & Mean ($\mu$) & Standard deviation $\sigma$ & Production seed value\\ \toprule
        Mean KS p-value across parameters & 0.0556 & 0.0074 & 0.0544\\ \midrule
        Max KS p-value for single parameter & 0.0852 & 0.0207 & 0.064\\ \midrule
        Mean coverage error across parameters & 0.0119 & 0.0025 & 0.014\\ \midrule
        Max coverage error for single parameter & 0.0207 & 0.0067 & 0.0256\\ \midrule
        Normalised posterior-mean RMSE & 0.087 & 0.0004 & 0.087\\ \midrule
        TARP ATC value & 0.0085 & 0.0068 & -0.0033 \\ \bottomrule
    \end{tabular}
    \caption{The calibration metrics used to evaluate the goodness of the trained flows for the 10 seeds of the best hyperparameters. Additionally the calibration metrics evaluated on the production seed, which is closest to the median of the 10 seeds. The  KS values are evaluated with 500 posterior samples, the TARP value is based on 300 samples and the coverage uses 5000 simulations with 500 posterior observations each. }
    \label{tab:calib-metrics}
\end{table*}

\begin{figure}
    \centering
    \includegraphics[width=1\linewidth]{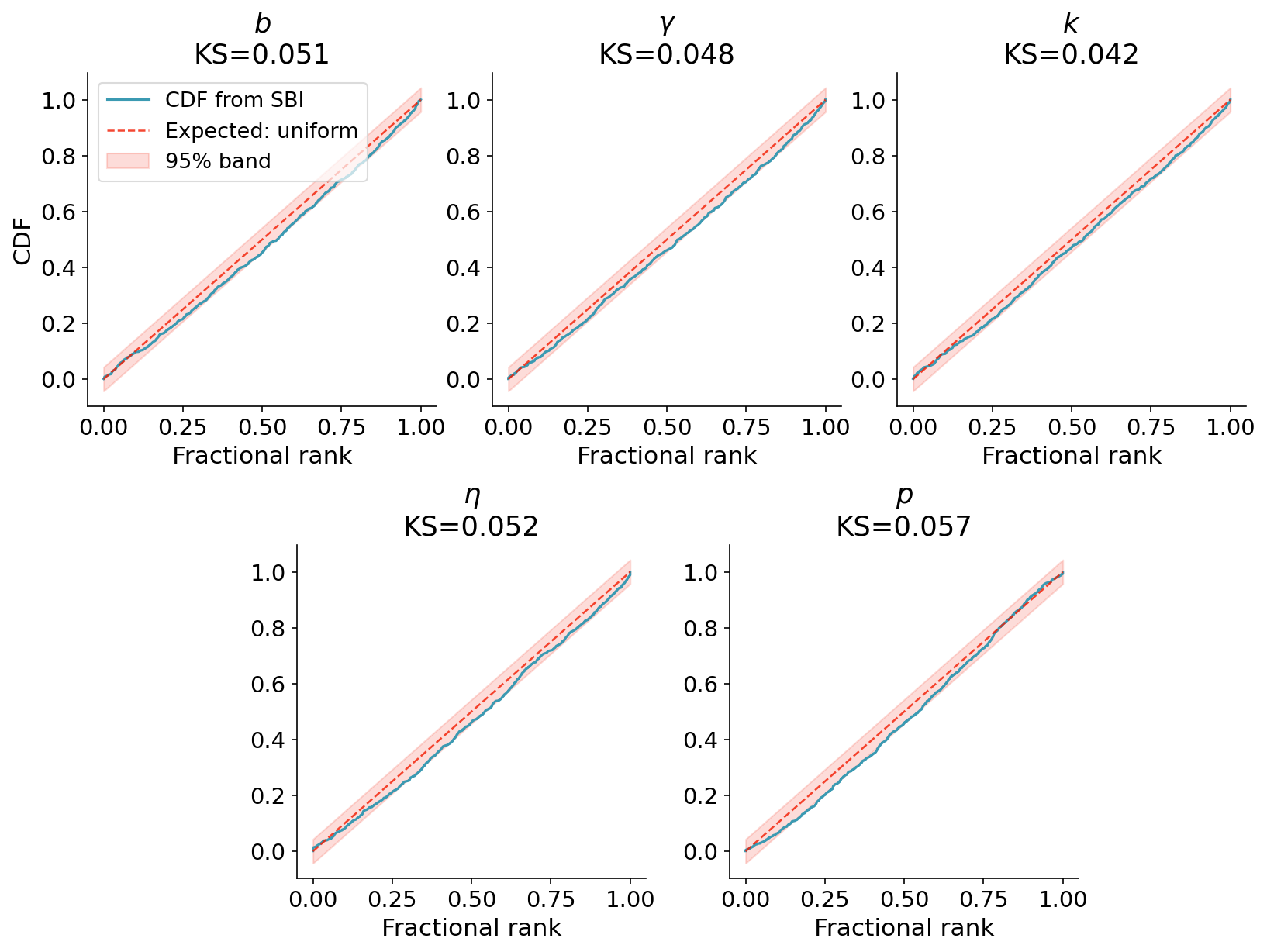}
    \caption{SBC rank plots for each parameter. Resulted from sampling 5000 posterior observations with 1000 samples each. The KS test for a uniform distribution is included as a dashed red reference line with a $95\%$ confidence interval in the shaded region.}
    \label{fig:sbc-rank-plots}
\end{figure}

\begin{figure}
    \centering
    \includegraphics[width=1\linewidth]{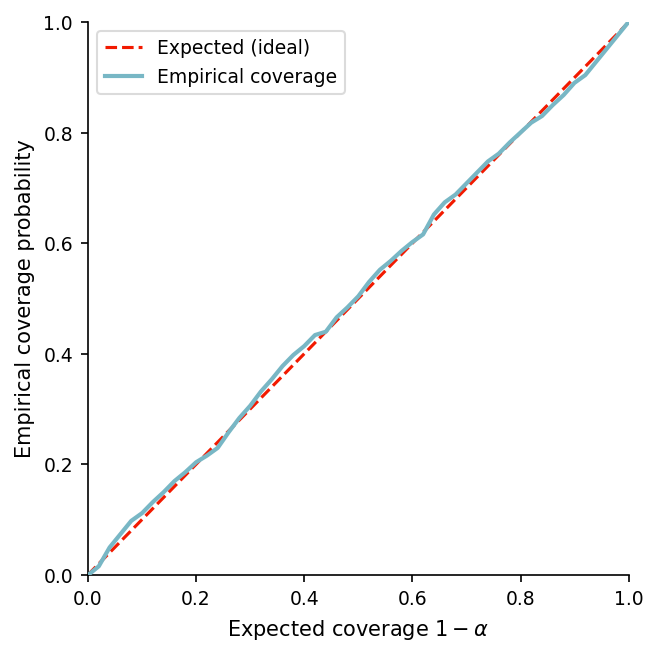}
    \caption{TARP analysis of the trained SBI flow, which displays global calibration of the flow.}
    \label{fig:tarp}
\end{figure}

\bsp	
\label{lastpage}
\end{document}